\documentclass{amsart}
\usepackage{natbib}
\usepackage{amsmath,amssymb,amsthm}
\usepackage{graphicx}
\usepackage{enumerate}         
\usepackage[section]{placeins} 

\newtheorem{thm}{Theorem}[section]
\newtheorem{cor}[thm]{Corollary}
\newtheorem{lem}[thm]{Lemma}
\newtheorem{prop}[thm]{Proposition}
\theoremstyle{definition}
\newtheorem{defn}[thm]{Definition}
\newtheorem{cond}[thm]{Condition}
\theoremstyle{remark}
\newtheorem{rem}[thm]{Remark}
\numberwithin{equation}{section}
\def\JELname{\textbf{Journal of Economic Literature Classification}\enspace}
\def\JEL#1{\par\addvspace\medskipamount{\rightskip=0pt plus1cm
\def\and{\ifhmode\unskip\nobreak\fi\ $\cdot$
}\noindent\JELname\ignorespaces#1\par}}

\newcommand{\set}[1]{\left\{#1\right\}}
\newcommand{\Ind}[1]{\mathbf{1}_{\left\{#1\right\}}}

\newcommand{\RR}{\mathbb{R}}

\newcommand{\CC}{\mathbb{C}}

\newcommand{\EE}{\mathbb{E}}

\newcommand{\cA}{\mathcal{A}}
\newcommand{\cB}{\mathcal{B}}

\newcommand{\cD}{\mathcal{D}}
\newcommand{\cF}{\mathcal{F}}

\newcommand{\cQ}{\mathcal{Q}}

\newcommand{\cU}{\mathcal{U}}

\renewcommand{\Re}{\mathrm{Re}}

\newcommand{\basymp}{b_{\mathrm{asymp}}}
\newcommand{\bnorm}{b_{\mathrm{norm}}}
\newcommand{\binv}{b_{\mathrm{inv}}}
\newcommand{\rt}{(r_t)_{t\geq0}}
\newcommand{\Rplus}{\mathbb{R}_{\geqslant 0}}

\begin{document}
\title[Yield Curve Shapes in Affine One-Factor Models]{Yield Curve Shapes and the Asymptotic Short Rate Distribution in Affine One-Factor Models}

\author{Martin Keller-Ressel}
\thanks{Supported by the Austrian Science Fund (FWF) through project P18022 and the START programm Y328.}
\address{Vienna University of Technology, Wiedner Hauptstrasse 8--10,
A-1040 Wien, Austria} \email{mkeller@fam.tuwien.ac.at}

\author{Thomas Steiner}
\thanks{Supported by the module M5 ``Modelling
of Fixed Income Markets'' of the PRisMa Lab, financed by Bank
Austria and the Republic of Austria through the Christian Doppler
Research Association.}
\address{Vienna University of Technology, Wiedner Hauptstrasse 8--10,
A-1040 Wien, Austria} \email{thomas@fam.tuwien.ac.at}

\thanks{Both authors would like to thank
Josef Teichmann for most valuable discussions and encouragement.
We also thank various proof-readers at FAM for their comments.}



\date{\today}

\begin{abstract}
We consider a model for interest rates, where the short rate is
given under the risk-neutral measure by a time-homogenous,
one-dimensional affine process in the sense of
\citeauthor*{Schachermayer}. We show that in such a model yield
curves can only be normal, inverse or humped (i.e. endowed with a
single local maximum). Each case can be characterized by simple
conditions on the present short rate $r_t$. We give conditions
under which the short rate process will converge to a limit
distribution and describe the risk-neutral limit distribution in
terms of its cumulant generating function. We apply our results to
the Vasi\v{c}ek model, the CIR model, a CIR model with added jumps
and a model of Ornstein-Uhlenbeck type. 
\end{abstract}

\maketitle

\section{Introduction}

 We consider a model for the term structure of interest rates, where the short
rate $\rt$ is given under the martingale measure by a
one-dimensional conservative affine process in the sense of
\citet*{Schachermayer}. An affine short rate process of this type
will lead to an exponentially-affine structure of zero-coupon bond
prices and thus also to an affine term
structure of yields and forward rates.\\
We emphasize here that the definition of \citet{Schachermayer} is
not limited to diffusions, but also includes processes with jumps
and even with jumps whose intensity depends in an affine way on
the state of the process itself. The class of models we consider
naturally includes the Vasi\v{c}ek model, the CIR model and
variants of them that are obtained by adding jumps, such as the
JCIR-model of \citet[Section~22.8]{BrigoMerc}. Since they are the
best-known, the two `classical' models of Vasi\v{c}ek and
Cox-Ingersoll-Ross will serve as the starting
point for our discussion of yield curve shapes:\\
A common criticism of the (time-homogenous) CIR and the
Vasi\v{c}ek model is that they are not flexible enough to
accommodate more complex shapes of yield curves, such as curves
with a dip (a local minimum), curves with a dip and a hump, or
other shapes that are frequently observed in the markets. Often
these shortcomings are explained by `too few parameters' in the
model (cf. \citet[Section~2.3.5]{Carmona} or
\citet[Section~3.2]{BrigoMerc}). However if jumps are added to the
mentioned models, additional parameters (potentially infinitely
many) are introduced through the jump part, while the model still
remains in the scope of affine models. It is not clear per se what
consequences the introduction of jumps will have for
the range of attainable yield curves, and this is one question we intend to answer in this article.\\
Moreover, there seems to be some confusion about what shapes of
yield curves are actually attainable even in well-studied models
like the CIR-model. While most sources (including the original
paper of \citet{CIR}) mention inverse, normal and humped shapes,
\citet[Section~2.3.5]{Carmona} write that \textit{`tweaking the
parameters [of the CIR model] can produce yield curves with one
hump or one dip'}, and \citet[Section~3.2]{BrigoMerc} state that
\textit{`some typical shapes, like that of an inverted yield
curve, may not be reproduced by the [CIR or Vasi\v{c}ek] model.'}
In our main result, Theorem \ref{Thm:yield_curves}, we settle this
question and prove that in any time-homogenous, affine one-factor
model the attainable yield curves are either inverse, normal or
humped. The proof will rely only on tools of elementary analysis
and on the characterization of affine processes through the
generalized Riccati equations of
\citet{Schachermayer}.\\
Another related problem is how the shape of the yield curve is
determined by the parameters of the model, and also how -- when
the parameters are fixed -- the yield curve is determined by the
level of the current short rate. We show in Section~\ref{Sec:CIR}
that also in this respect the CIR model has not been completely
understood and discuss a misconception that originates in
\citep{CIR} and is repeated for example in \citep{Rebonato}.\\

In Section \ref{Sec:limit} we provide conditions under which an
affine process converges to a limit distribution. We also
characterize the limit distribution in terms of its cumulant
generating function, extending results of \citet{JurekVervaat} and
\citet{SatoYamazato} for OU-type processes to the class of affine
processes. These results can again be interpreted in the context
of interest rates, where they can be used to derive the
risk-neutral asymptotic distribution of the short rate $\rt$ as
$t$ goes to
infinity.\\

We conclude our article in Section \ref{Sec:applications} by
applying the theoretical results to several interest rate models,
such as the Vasi\v{c}ek model, the CIR model, the JCIR model and
an Ornstein-Uhlenbeck-type model.

\section{Preliminaries}\label{Sec:preliminaries}
In this section we collect some key results on affine processes
from \citet{Schachermayer}. In their article affine processes are
defined on the $(m+n)$-dimensional state space $\Rplus^m \times
\RR^n$, and we will try to simplify notation where this is
possible in the one-dimensional case. Results on affine processes
with state space $\Rplus$ can also be found in \citet{Filipovic}.

\begin{defn}[One-dimensional affine process]\label{Def:Affine_proc}
A time-homogenous Markov process $\rt$ with state space $D =
\Rplus$ or $\RR$ and its semi-group $(P_t)_{t \ge 0}$ are called
\emph{affine}, if the characteristic function of its transition
kernel $p_t(x,.)$, given by
\[\widehat{p}_t(x,u) = \int_D{e^{u\xi}\;p_t(x,d\xi)}\]
and defined (at least) on
\begin{equation*}\label{Eq:U_defn}
    \cU = \begin{cases}\set{u \in \CC : \Re\,u \leq 0} &\quad \text{if} \quad D = \Rplus\;,\\ \set{u \in \CC : \Re\,u = 0} &\quad \text{if} \quad D = \RR\;, \end{cases}
\end{equation*}
is \emph{exponentially affine} in $x$. That is, there exist
$\CC$-valued functions $\phi(t,u)$ and $\psi(t,u)$, defined on
$\Rplus \times \cU$, such that
\begin{equation}\label{Eq:trans_kernel_affine}
\widehat{p}_t(x,u) = \exp\left(\phi(t,u) + x \psi(t,u)\right)
\quad \text{for all} \quad x \in D, (t,u) \in \Rplus \times \cU\;.
\end{equation}
\end{defn}

For subsequent results the following regularity condition for
$\rt$ will be needed:

\begin{defn}An affine process is called \emph{regular} if it is
stochastically continuous and the right hand derivatives
\[\partial_t^+\,\phi(t,u)|_{t=0} \qquad \text{and} \qquad \partial_t^+\,\psi(t,u)|_{t=0}\]
exist for all $u \in \cU$ and are continuous at $u = 0$.
\end{defn}

\begin{defn}\label{Def:Admissible}
The parameters $(a, \alpha, b, \beta, c, \gamma, m, \mu)$ are
called \emph{admissible} for a process with state space $\Rplus$
if
\begin{gather*}
a = 0,\\
\alpha, b, c, \gamma \in \Rplus\,,\\
\beta \in \RR\,,\\
\begin{split}m, \mu \; &\text{are L\'{e}vy measures on $(0,\infty)$, where
$m$ satisfies}\\ &\int_{(0,\infty)}{(\xi \land 1)\,m(d\xi)} <
\infty\;,
\end{split}
\end{gather*}
and admissible for a process with state space $\RR$ if
\begin{gather*}
a,c \in \Rplus\,,\\
b,\beta \in \RR\,,\\
m \;\text{is a L\'{e}vy measure on} \; \RR \setminus \set{0}\;,\\
\alpha = 0, \gamma = 0, \mu \equiv 0\;.
\end{gather*}
Moreover define the truncation functions
\begin{equation*}
h_F(\xi) = \begin{cases}0\quad &\text{if} \quad
D=\Rplus\\\frac{\xi}{1+\xi^2}\quad &\text{if} \quad D = \RR
\end{cases} \qquad \text{and} \qquad
h_R(\xi) = \begin{cases}\frac{\xi}{1+\xi^2}\quad &\text{if} \quad
D=\Rplus\\0 \quad &\text{if} \quad D = \RR
\end{cases},
\end{equation*}
and finally the functions $F(u)$, $R(u)$ for $u \in \CC$ as
\begin{gather}
F(u) = au^2 + bu - c + \int_{D \setminus \set{0}}{\left(e^{u\xi} -
1 -
u h_F(\xi)\right)\,m(d\xi)}\label{Eq:F_def}\;,\\
R(u) = \alpha u^2 + \beta u - \gamma + \int_{D \setminus
\set{0}}{\left(e^{u\xi} - 1 - u
h_R(\xi)\right)\,\mu(d\xi)}\;.\label{Eq:R_def}
\end{gather}
\end{defn}

The next result is a one-dimensional version of the key result of
\citet{Schachermayer}:

\begin{thm}[\citeauthor*{Schachermayer}, Theorem~2.7]\label{Thm:Schachermayer_1}
Suppose $\rt$ is a one-dimensional regular affine process. Then it
is a Feller process. Let $\cA$ be its infinitesimal generator.
Then $C_c^\infty(D)$ is a core of $\cA$, $C_c^2(D) \subseteq
\cD(\cA)$ and there exist some admissible parameters $(a, \alpha,
b, \beta, c, \gamma, m, \mu)$ such that, for $f \in C_c^2(D)$,
\begin{align}
\cA f(x) &= (a + \alpha x) f''(x) + (b + \beta x) f'(x) - (c + \gamma x)f(x) + \notag \\
 &+ \int_{D
\setminus \set{0}}{\left(f(x + \xi) - f(x) -
f'(x)h_F(\xi)\right)\,m(d \xi)} + \notag
\\ &+ x \int_{D \setminus \set{0}}{\left(f(x + \xi) - f(x) -
f'(x)h_R(\xi)\right)\, \mu(d \xi)}\;. \label{Eq:generator_affine}
\end{align}
Moreover $\phi(t,u)$ and $\psi(t,u)$, defined by
\eqref{Eq:trans_kernel_affine}, solve the \emph{generalized
Riccati equations}
\begin{subequations}\label{Eq:Riccati_general}
\begin{align}
\partial_t\,\phi(t,u) &= F\left(\psi(t,u)\right), \qquad \phi(0,u)
= 0\;, \label{Eq:Riccati_general_1}\\
\partial_t\,\psi(t,u) &= R\left(\psi(t,u)\right), \qquad \psi(0,u)
= u\;. \label{Eq:Riccati_general_2}
\end{align}
\end{subequations}
Conversely let $(a, \alpha, b, \beta, c, \gamma, \mu, m)$ be some
admissible parameters. Then there exists a unique regular affine
semigroup $(P_t)_{t \ge 0}$ with infinitesimal generator
\eqref{Eq:generator_affine}, and \eqref{Eq:trans_kernel_affine}
holds with $\phi(t,u)$ and $\psi(t,u)$ given by
\eqref{Eq:Riccati_general}.
\end{thm}

Closely related to affine processes is the notion of an
Ornstein-Uhlenbeck (OU-)type process. These processes are of some
importance, since they usually offer good analytic tractability
and have been studied for longer than affine processes. Following
\citet[Chapter~17]{Sato} an OU-type process $(X_t)_{t\ge 0}$ can
be defined as the solution of the Langevin SDE
\[dX_t = -\lambda X_t\,dt + dL_t, \qquad \lambda \in \RR, X_0 \in \RR,\]
where $(L_t)_{t \ge 0}$ is a L\'{e}vy process, often called
background driving L\'{e}vy process (BDLP). In an equivalent
definition, an OU-type process is a time-homogenous Markov
process, whose transition kernel $p_t(x,.)$ has the characteristic
function
\[\widehat{p}_t(x,u) = \exp\left(\int_0^t{F(e^{-\lambda s}u)\,ds} + x e^{-\lambda t} u\right)\;,\]
where $F(u)$ is the characteristic exponent of $(L_t)_{t \ge 0}$.
From the last equation it is immediately seen that every OU-type
process is an affine process in the sense of Definition
\ref{Def:Affine_proc}. It is also seen that in the generalized
Riccati equations \eqref{Eq:Riccati_general} for an OU-type
process necessarily $R(u) = -\lambda u$. Comparing this with
\eqref{Eq:R_def} and Definition \ref{Def:Admissible}, it is seen
that \emph{any} regular affine process with state space $\RR$ is a
process of OU-type. The reverse, however is not true, as there
also exist OU-type processes with state space $\Rplus$. We will
give an example of such a process in Section \ref{Sec:Gamma}.\\

Naturally we will not only be interested in the process $\rt$
itself, but also in its integral $\int_0^t{r_s\,ds}$ and in
quantities of the type
\begin{equation}\label{Eq:pricing_semigroup}
\cQ_t\,f(x) :=
\EE\left[\left.\exp\left(-\int_0^t{r_s\,ds}\right)f(r_t)\right|r_0
= x\right]\;,
\end{equation}
where $f$ is a bounded function on $D$. The next result is an
application of the Feynman-Kac formula for Feller semigroups (cf.
\citet[Section III.19]{RogersWilliams}) and can be found in
\citet{Schachermayer}. It relies on the positivity of $\rt$ and is
therefore only applicable if $D = \Rplus$.

\begin{prop}[\citeauthor*{Schachermayer}, Proposition~11.1]\label{Prop:generator_shifted}
Let $\rt$ be a one-dimensional, regular affine process with state
space $\Rplus$. Then the family $(\cQ_t)_{t \geq 0}$ defined by
\eqref{Eq:pricing_semigroup} forms a regular, affine semigroup
with infinitesimal generator
\[\cB f(x) = \cA f(x) - xf(x) \quad \text{for all} \quad f \in C_c^2(D)\;.\]
\end{prop}

We will make extensive use of the convexity and continuous
differentiability of the functions $F$ and $R$ from
Definition~\ref{Def:Admissible}. These properties are established
in this Lemma:

\begin{lem}\label{Lem:FR_properties}
If $c=\gamma=0$ then $F,\,R$ as defined in
Definition~\ref{Def:Admissible} have the following properties:
\begin{enumerate}[(i)]
    \item $R(0) = 0$ and $F(0) = 0$.
    \item $R(u) < \infty$ for all $u \in (-\infty,0]$.
    \item If $F(u) < \infty$ on $(c_1,c_2) \subseteq \RR$, then $F$ is either strictly convex on $(c_1,c_2)$ or
    $F(u) = bu$ for all $u \in \RR$. The same holds for $R$ with $b$ replaced by $\beta$.
    \item If $F(u) < \infty$ on $(c_1,c_2) \subseteq \RR$, then $F$ is continuously differentiable on
    $(c_1,c_2)$. Also the one-sided derivatives at $c_1$ and $c_2$ are defined but may take the
    values $-\infty$ (at $c_1$) and $+\infty$ (at $c_2$).
    The same holds for $R$.
\end{enumerate}
\end{lem}
\begin{proof}
Property (i) is obvious. If $D = \RR$ then by
Definition~\ref{Def:Admissible} $R(u) = \beta u$ such that (ii)
follows immediately. If $D = \Rplus$ we use the estimate
\begin{equation}\label{Eq:estimate_R} |e^{u\xi}-1-uh_R(\xi)| \le
|u|\left(\mathcal{O}(\xi^2) \wedge 1\right)\;,
\end{equation}
for all $u \in (-\infty,0]$ and $\xi \in \Rplus$, and (ii) follows
from \eqref{Eq:R_def}. For Property (iii) note that by the
L\'{e}vy-Khintchine formula there exists an infinitely divisible
random variable $X$, such that $F$ is its cumulant generating
function, i.e. $F(u) = \log \EE\left[e^{uX}\right]$ for $u \in
(c_1,c_2)$.
Choosing two distinct numbers $u,v \in (c_1,c_2)$, we apply the
Cauchy-Schwarz inequality to
\[F\left(\frac{u+v}{2}\right) = \log \EE\left[e^{\frac{uX}{2}}\cdot e^{\frac{vX}{2}}\right] \leq
 \log \sqrt{\EE[e^{uX}] \cdot \EE[e^{vX}]} = \frac{F(u) + F(v)}{2}\]
 which shows convexity of $F$.
The inequality is strict unless there exists some $c \neq 0$ such
that $e^{uX} = c e^{vX}$ almost surely. This can only be the case
if $X$ is constant a.s., in which case $F$ is linear. The same
argument applies to $R$. Property (iv) follows from the convexity
and from the fact that $F$ and $R$ are analytic on $\set{u \in \CC
: \Re\,u \in (c_1,c_2)}$ (cf. \citet[Chapter~7]{Lukacs}).
\end{proof}

\section{Theoretical Results}
\label{Sec:theory}

We will now use the theory from the last section to calculate bond
prices, yields and other quantities in an interest rate model
where the short rate follows a one-dimensional regular affine
process $\rt$ under the martingale measure. Naturally we will also
make the assumption that $\rt$ is conservative, i.e. that
$p_t(x,D) = 1$ for all $(t,x) \subseteq \Rplus \times D$. This
implies by \citet[Proposition~9.1]{Schachermayer} that $c = \gamma
= 0$ in Definition~\ref{Def:Admissible}. We will need some
additional assumptions which are summarized in the following
condition:

\begin{cond}\label{Cond:rt}
The one-dimensional affine process $\rt$ is assumed to be regular
and conservative. In addition, if the process has state space
$\RR$, such that by Definition~\ref{Def:Admissible} $R(u) = \beta
u$, we require that
\begin{equation}\label{Eq:F_cond}
F(u) < \infty \quad \text{for all} \quad  u \in
\begin{cases}(1/\beta,0] &\quad \text{if} \quad \beta < 0\;,\\
(-\infty,0] &\quad \text{else}\;.\end{cases}
\end{equation}
\end{cond}

It will be seen that the condition on $F$ is necessary to
guarantee existence of bond prices for all maturities in the term
structure model. By \citet[Theorem~25.17]{Sato} we get an
equivalent formulation of Condition \ref{Cond:rt}, if we replace
$F(u) < \infty$ by $\int_{|\xi|
> 1}{e^{u\xi}\,m(d\xi)} < \infty$. Next we define a quantity that
will generalize the coefficient of mean reversion from OU-type
processes:

\begin{defn}[quasi-mean-reversion]\label{Def:quasi_mean_reversion}
Given a one-dimensional conservative affine process $\rt$, define
the \emph{quasi-mean-reversion} $\lambda$ as the positive solution
of
\begin{equation}\label{Eq:define_p}
R(-1/\lambda) = 1\;.
\end{equation}
If there is no positive solution we set $\lambda = 0$.
\end{defn}

Since $R$ is by Lemma~\ref{Lem:FR_properties} a convex function
satisfying $R(0) = 0$, it is easy to see that \eqref{Eq:define_p}
can have at most one solution and thus $\lambda$ is well-defined.
The name \emph{quasi-mean-reversion} is derived from the fact that
if $\rt$ is a process of OU-type with positive mean-reversion,
then $R(u) = \beta u$ and the quasi-mean-reversion $\lambda =
-\beta$ is exactly the coefficient of mean reversion of $\rt$.
When the process $\rt$ satisfies Condition \ref{Cond:rt}, it is
seen that $F$ must be defined at least on
$(-1/\lambda,0]$.\\
We will encounter several times the condition that $\lambda > 0$.
The next result gives an equivalent formulation in terms of
$(\alpha, \beta, \mu)$:

\begin{prop} The quasi-mean reversion $\lambda$ is strictly positive if and
only if $\alpha > 0$, $\int_{D \setminus
\set{0}}{h_R(\xi)\,\mu(d\xi)} = \infty$, or $\beta - \int_{D
\setminus \set{0}}{h_R(\xi)\,\mu(d\xi)} < 0$.
\end{prop}
\begin{proof}
First note that by Lemma~\ref{Lem:FR_properties} $R(u) < \infty$
for all $u \in (-\infty,0]$. Using the estimate
\eqref{Eq:estimate_R} and a dominated convergence argument it is
seen from \eqref{Eq:R_def} that
\begin{align}
&\lim_{u \to -\infty}\frac{R(u)}{u^2} = \alpha \label{Eq:R_limit1}\\
&\lim_{u \to -\infty}\frac{R(u) - \alpha u^2}{u} = \beta_0 :=
\beta - \int_{D \setminus \set{0}}{h_R(\xi)\,\mu(d\xi)}\;,
\label{Eq:R_limit2}
\end{align}
where $\beta_0$ can also take the value $-\infty$. Suppose now
that $\alpha > 0$. Then by \eqref{Eq:R_limit1} we get $\lim_{u \to
-\infty} R(u) = \infty$. Since $R(0) = 0$ and $R$ is continuous it
follows that there exists a $\lambda > 0$ such that $R(-1/\lambda)
= 1$. Similarly if $\alpha = 0$, but $\beta_0 < 0$, it follows
from \eqref{Eq:R_limit2} that $\lim_{u \to -\infty}R(u)
= \infty$ and thus again that $\lambda > 0$.\\
Conversely, suppose that $\alpha = 0$ and $\beta_0 \ge 0$. Then
\[\lim_{u \to -\infty}R'(u) = \lim_{u \to -\infty}\frac{R(u)}{u} =
\beta_0 \ge 0\;.\] By the convexity of $R$ it follows that $R'(u)
\ge 0$ for all $u \in (-\infty,0)$. Since $R(0) = 0$ this implies
that $R(u) \le 0$ for all $u \in (-\infty,0)$, and consequently
that $\lambda = 0$.
\end{proof}

\subsection{Bond Prices}
We consider now the price $P(t,t+x)$ of a zero-coupon bond with
time to maturity $x$, at time $t$,  given by
\[P(t,t+x) =
\EE\left[\left.\exp\left(-\int_t^{t+x}{r_s\,ds}\right)\right|\cF_t\right]\;.\]
The affine structure of $\rt$ carries over to the bond prices, and
we get the following result:

\begin{prop}\label{Prop:bond_prices}
Let the short rate be given by a one-dimensional affine
process $\rt$ satisfying Condition \ref{Cond:rt}.\\
Then the bond price $P(t,t+x)$ exists for all $t,x \geq 0$ and is
given by
\begin{equation}\label{Eq:affine_bond_price}
P(t,t+x) = \exp\left(A(x) + r_t B(x)\right)
\end{equation}
where $A$ and $B$ solve the generalized Riccati equations
\begin{subequations}\label{Eq:Riccati_bondprice}
\begin{align}
\partial_x A(x) &= F\left(B(x)\right) \qquad &A(0) = 0\;, \label{Eq:Riccati_bondprice1}\\
\partial_x B(x) &= R\left(B(x)\right) - 1 \qquad &B(0) = 0\;. \label{Eq:Riccati_bondprice2}
\end{align}
\end{subequations}
\end{prop}

\begin{proof}
If $D = \Rplus$ the assertion follows directly from
Proposition~\ref{Prop:generator_shifted} by noting that $P(t,t+x)
=
\cQ_x\,1$.\\
If $D = \RR$ then, as discussed after
Theorem~\ref{Thm:Schachermayer_1}, $\rt$ is a process of OU-type
and $R(u)$ has the simple structure $R(u) = \beta u$. By
\citet[(17.2) - (17.3)]{Sato} we obtain in this case directly that
\begin{equation}\label{Eq:bond_interm}
\EE\left[\exp\left(-\int_t^{t +x}{r_s\,ds}\right)\right] =
\exp\left(\int_0^x{F(B(s))\,ds} + r_t B(x)\right)\;,
\end{equation}
with $B(x) = (1 - e^{\beta x})/\beta$ if $\beta \neq 0$ and $B(x)
= -x$ when $\beta = 0$. As a function of $x \in \Rplus$, $B$ is
continuously decreasing from $0$ to $1/\beta$ if $\beta < 0$, and
from $0$ to $-\infty$ if $\beta \ge 0$. It is therefore seen that
the integral on the right side of \eqref{Eq:bond_interm} is finite
for all $x \in \Rplus$ if and only if $F$ satisfies
\eqref{Eq:F_cond}, as required by Condition \ref{Cond:rt}.
\end{proof}

\begin{cor}\label{Cor:B_properties}
Let $\rt$ satisfy Condition \ref{Cond:rt} and have
quasi-mean-reversion $\lambda$. Then the function $B(x)$ from
Proposition~\ref{Prop:bond_prices} is strictly decreasing and
satisfies
\[\lim_{x \to \infty}B(x) = -1/\lambda\;.\]
\end{cor}
\begin{proof}
The result follows from a qualitative analysis of the autonomous
ODE \eqref{Eq:Riccati_bondprice2}. Let $\lambda > 0$. Since
$R(-1/\lambda) - 1 = 0$ the point $x_* := -1/\lambda$ is an
critical point of \eqref{Eq:Riccati_bondprice2}. By the convexity
of $R$ and the fact that $R(0) = 0$ it follows that $R'(x_*) < 0$
such that $x_*$ is asymptotically stable, i.e. solutions entering
a small enough neighborhood of $x_*$ must converge to $x_*$. Since
$R(x) - 1 < 0$ for $x \in (x_*,0]$ and
 there is no other critical point in
$(x_*,0]$, we conclude that $B(x)$ -- the solution of \eqref{Eq:Riccati_bondprice2} starting at $0$ -- is strictly decreasing and converges to $x_*$.\\
If $\lambda = 0$ then there is no critical point in $(-\infty,0]$
and $R(x) - 1 < 0$ for $x \in (-\infty,0]$. It follows that $B(x)$
is strictly decreasing and diverges to $-\infty$.
\end{proof}

\subsection{The Yield Curve and the Forward Rate Curve}

The next results are the central theoretical results of this
article and describe the global shapes of attainable yield curves
in any affine one-factor term structure model.
\begin{defn}
The (zero-coupon) yield $Y(r_t,x)$ is given by $Y(r_t,0) := r_t$
and
\begin{equation}\label{Eq:yield_def}
Y(r_t,x) := -\frac{\log P(t,t+x)}{x} = - \frac{A(x)}{x} - r_t
\frac{B(x)}{x} \quad \text{for all}\quad x > 0\;.
\end{equation}
For $r_t$ fixed, we call the function $Y(r_t,.)$ the
\textbf{yield curve}.\\
The (instantaneous) forward rate $f(r_t,x)$ is given by $f(r_t,0)
:= r_t$ and
\begin{equation}\label{Eq:fwd_def}
f(r_t,x) := - \partial_x\, \log P(t,t+x) = - A'(x) - r_t B'(x)
\quad \text{for all} \quad x > 0\;.
\end{equation}
For $r_t$ fixed, we call the function $f(r_t,.)$ the
\textbf{forward rate curve}.
\end{defn}
By l'Hospital's rule and the generalized Riccati equations
\eqref{Eq:Riccati_bondprice} it is
seen that both the yield and the forward rate curve are continuous at $0$.\\
The first quantity associated to the yield curve that we consider,
is the asymptotic level $\basymp$ of the yield curve as $x \to
\infty$, also known as long-term yield, consol yield or simply
`long end'.

\begin{thm}\label{Thm:basymp}
Let the short rate process be given by a one-dimensional affine
process $\rt$ satisfying Condition \ref{Cond:rt} with
quasi-mean-reversion $\lambda$.\linebreak If $\lambda > 0$ then
\[\basymp := \lim_{x \to \infty}Y(r_t,x) = \lim_{x \to \infty}f(r_t,x) = -F(-1/\lambda)\;.\]
If $\lambda = 0$ then
\[\basymp = \lim_{u \to -\infty} -F(u) + r_t \left(1 - R(u)\right)\;.\]
\end{thm}

\begin{proof}
From \eqref{Eq:Riccati_bondprice1} we obtain that
\begin{equation}\label{Eq:basymp_interm1}
\lim_{x \to \infty}\frac{A(x)}{x} = \lim_{x \to \infty}A'(x) =
\lim_{x \to \infty}F(B(x))\;.
\end{equation}
If $\lambda > 0$ then by Corollary \ref{Cor:B_properties}
\begin{equation}\label{Eq:basymp_interm2}
\lim_{x \to \infty} B(x) = -1/\lambda, \quad \lim_{x \to
\infty}{\frac{B(x)}{x}} = 0 \quad \text{and} \quad \lim_{x \to
\infty}B'(x) = 0
\end{equation}
and the assertion follows by combining \eqref{Eq:yield_def}~--~\eqref{Eq:basymp_interm2}.\\
If $\lambda = 0$ then $\lim_{x \to \infty}{B(x)} = -\infty$ and
\[\lim_{x \to \infty}{\frac{B(x)}{x}} = \lim_{x \to \infty}B'(x) = \lim_{x \to \infty}R(B(x)) -
1\;.\] By setting $u := B(x)$ we obtain the desired result.
\end{proof}

From Theorem~\ref{Thm:basymp} it is clear that for practical
purposes only models with $\lambda > 0$ will be useful. So far we
know that in this case the short end of the yield curve is given
by $Y(r_t,0) = r_t$ and the long end by $Y(r_t,\infty) = \basymp$.
We will now examine what happens between these two endpoints.

\begin{defn}\label{Defn:classify}
The yield curve $Y(r_t,x)$ is called
\begin{itemize}
    \item \textbf{normal} if it is a strictly increasing function of $x$,
    \item \textbf{inverse} if it is a strictly decreasing function of $x$,
    \item \textbf{humped} if it has exactly one local maximum and no
    minimum on $(0,\infty)$.
\end{itemize}
In addition we call the yield curve \textbf{flat} if it is
constant over all $x \in \Rplus$.
\end{defn}

This is our main result on the shapes of yield curves in affine
one-factor models:

\begin{thm}\label{Thm:yield_curves}
Let the risk-neutral short rate process be given by a
one-dimensional affine process $\rt$ satisfying Condition
\ref{Cond:rt} and with quasi-mean-reversion $\lambda > 0$. In
addition suppose that $F \neq 0$ and that either $F$ or $R$ is
non-linear. Then the following holds:
\begin{itemize}
    \item The yield curve $Y(r_t,.)$ can only be normal, inverse or
    humped.
    \item Define
\begin{equation*}\label{Eq:binv_bnorm}
    \bnorm := -\frac{F'(-1/\lambda)}{R'(-1/\lambda)} \qquad \text{and} \qquad \binv := \begin{cases}-\dfrac{F'(0)}{R'(0)}\; &\text{if} \; R'(0) < 0\\
    +\infty \; &\text{if} \; R'(0) \geq 0\;.\end{cases}
\end{equation*}
    The yield curve is normal if $\;r_t \leq \bnorm\,$, humped
    if $\;\bnorm < r_t < \binv\,$ and inverse if $\;r_t \geq
    \binv\,$.
\end{itemize}
\end{thm}

\begin{figure}
  \includegraphics[width=\textwidth]{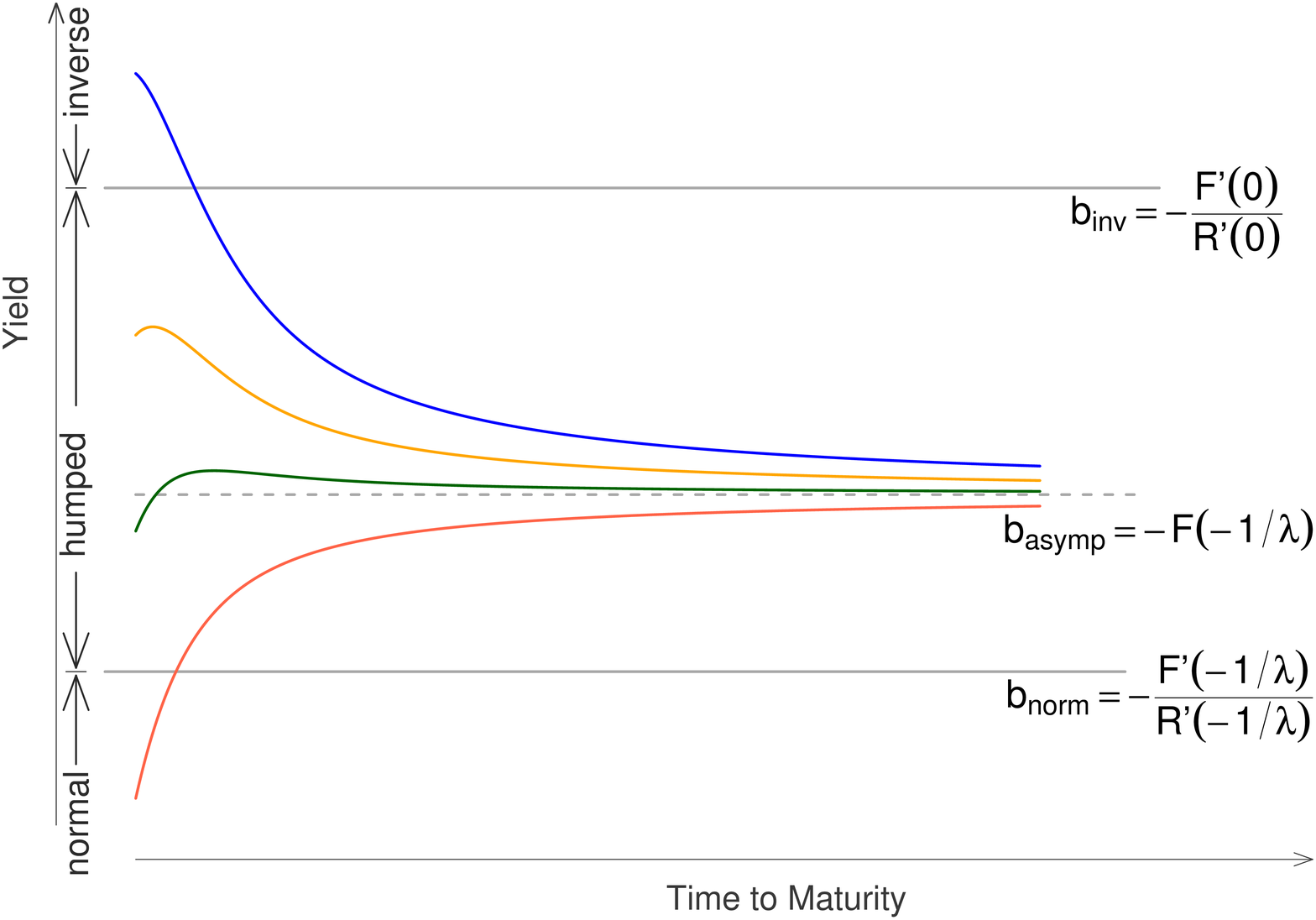}
  \caption{This Figure shows a graphical summary of Theorems \ref{Thm:basymp} and \ref{Thm:yield_curves}, as well as the definitions of the key
  quantities $\bnorm$, $\basymp$ and $\binv$. In \emph{any} affine model satisfying the conditions of Theorem \ref{Thm:yield_curves}, the shapes of
  yield curves will follow the picture given here. They will be normal if $r_0$ is below $\bnorm$, humped if $r_0$ is between $\bnorm$ and $\binv$ and inverse if $r_0$ is above $\binv$.
  Also all yield curves will tend asymptotically to the same level $\basymp$.}
  \label{Fig:yield}
\end{figure}

The above theorem is visualized in Figure~\ref{Fig:yield}. For its
proof we will use a simple Lemma. We state the Lemma without
proof, since it follows in an elementary way from the usual
definition of a convex function on $\RR$.
\begin{lem}\label{Lem:convex}A strictly convex or a strictly
concave function on $\RR$ intersects an affine function in at most
two points. In the case of two intersection points $p_1 < p_2$,
the convex function lies strictly below the affine function on the
interval $(p_1,p_2)$; if the function is concave it lies strictly
above the affine function on $(p_1,p_2)$.
\end{lem}

\begin{proof}[Proof of
Theorem~\ref{Thm:yield_curves}]
 Define the
function $H(x) \colon \Rplus \to \RR$ by
\begin{equation}\label{Eq:H_def}
H(x) := Y(r_t, x)x = -A(x) - r_t B(x)\;.
\end{equation}
We will see that the convexity behavior of $H$ will be crucial for
the shape of the yield curve $Y(r_t,.)$. From the generalized
Riccati equations \eqref{Eq:Riccati_bondprice} the first
derivative of $H$ is calculated as
\begin{equation}\label{Eq:Dx}
\partial_x H(x) = - F(B(x)) - r_t \left(R(B(x)) -
1\right)
\end{equation} and the second as
\begin{equation}\label{Eq:Dxx}
\partial_{xx}H(x) = -B'(x) \left(F'(B(x)) + r_t
R'(B(x))\right)\;.
\end{equation}
Note that $F$ and $R$ are continuously differentiable by
Lemma~\ref{Lem:FR_properties}, and also $B$ by
\eqref{Eq:Riccati_bondprice2}, such that the second derivative of
$H$ is well-defined and continuous. Since $B$ is strictly
decreasing by Corollary~\ref{Cor:B_properties}, the factor
$-B'(x)$ is positive for all $x \in \Rplus$. The sign of
$\partial_{xx}H(x)$ therefore equals the sign of
\begin{equation}\label{Eq:kx_definition}
k(x) := F'(B(x)) + r_t R'(B(x))\;.
\end{equation} From the fact that $B$ is decreasing and $F$ and
$R$ are convex it is obvious that $k$ must be decreasing. We will
now show that $k$ has at most a single zero in $[0,\infty)$:
\begin{enumerate}[(a)]
    \item $D = \Rplus$: We have assumed that either $F$ or $R$ is
    non-linear. By Lemma~\ref{Lem:FR_properties} this implies that
    either $F$ or $R$ is strictly convex, and thus that either
    $F'$ or $R'$ is strictly increasing. If $r_t > 0$, then it
    follows that $k$ is strictly decreasing and thus has at most a
    single zero. If $r_t = 0$, an additional argument is needed:
    It could happen that $F$ is of the form $F = bu$ such that
    $k(x) = b$ and $k$ is no longer strictly decreasing. However, by
    assumption, $F \neq 0$ such that in this case $k$ has no zero
    in $[0,\infty)$.
    \item $D = \RR$: In this case, by the admissibility conditions in
    Definition \ref{Def:Admissible}, we have necessarily $R(u) =
    \beta u$. Also, since either $F$ or $R$ is non-linear, $F$
    must be non-linear and thus by Lemma \ref{Lem:FR_properties}
    strictly convex. It follows that
    $k(x) = F'(B(x)) + r_t \beta$
    is strictly decreasing and thus has at most a single zero in
    $[0,\infty)$.
\end{enumerate}
We have shown that $k$ is decreasing and has at most a single
zero; to determine whether it has a zero for some value of $r_t$,
we consider the two `endpoints' $k(0)$ and $\lim_{x \to
\infty}k(x)$. First we show that
\begin{equation}
k(0) \geq 0 \quad \text{if and only if} \quad r_t \leq
\binv := \begin{cases}-\dfrac{F'(0)}{R'(0)}\; &\text{if} \; R'(0) < 0\\
    +\infty \; &\text{if} \; R'(0) \geq 0\;.\end{cases}\label{Eq:k_behaviour1}
\end{equation}
Since $B(0) = 0$ by Proposition~\ref{Prop:bond_prices} it follows
that
\[k(0) = F'(0) + r_t R'(0)\;.\]
We distinguish two cases:
\begin{enumerate}[(a)]
\item If $R'(0) < 0$ then the assertion \eqref{Eq:k_behaviour1}
follows immediately. \item Consider the case that $R'(0) \ge 0$:
Assume that $D = \RR$. Then we have $R(u) = \beta u$ and $R'(0) =
\beta \ge 0$. This, however, stands in contradiction to our
assumption $\lambda
> 0$, which implies that $\beta = -\lambda < 0$ (cf. Definition \ref{Def:quasi_mean_reversion}).
Thus we must have $D = \Rplus$ and $r_t \ge 0$; in this case it
follows that $k(0) \ge 0$, \emph{for all} $r_t \in D$, and we set
$\binv = +\infty$.
\end{enumerate}
Next we consider the right end of $k(x)$ and show that
\begin{equation}\label{Eq:k_behaviour2}
\lim_{x \to \infty}k(x) \leq 0 \quad \text{if and only if} \quad
r_t \geq \bnorm := -\frac{F'(-1/\lambda)}{R'(-1/\lambda)}\;.
\end{equation} Since $\lim_{x \to \infty}B(x) = -1/\lambda$ by
Corollary~\ref{Cor:B_properties} we have that
\begin{equation}\label{Eq_kinfty}
\lim_{x \to \infty}k(x) = F'(-1/\lambda) + r_t R'(-1/\lambda)\;.
\end{equation}
By assumption $\lambda > 0$, and by
Definition~\ref{Def:quasi_mean_reversion} it holds that
$R(-1/\lambda) = 1$. Also $R(0) = 0$, and by the mean value
theorem
\[1 = R(-1/\lambda) - R(0) = -\frac{1}{\lambda} R'(\xi)\]
for some $\xi \in (-1/\lambda,0)$. Since $R'$ is increasing, it
follows that $R'(-1/\lambda) \le -\lambda < 0$, and we can
deduce \eqref{Eq:k_behaviour2} directly from \eqref{Eq_kinfty}.\\
We summarize our results on the function $k$ so far: $k$ stays
negative on $(0,\infty)$ if $r_t \geq \binv$ and positive if $r_t
\leq \bnorm$. It has a single zero on $(0,\infty)$ if and only if
$\bnorm < r_t < \binv$. If $k$ has a zero on $(0,\infty)$, since
$k$ is decreasing, the sign of $k$ will be positive to the left of
the zero and negative to the right of the zero.\\

Since $\partial_{xx}H$ has the same sign as $k$, the statements
above translate in the obvious way to the convexity behavior of
$H$. We will now use the convexity behavior of $H$ to derive our
results
about the yield curve.\\
Consider the equation
\begin{equation}\label{Eq:H_affine}
H(x) = cx, \qquad x \in [0,\infty)
\end{equation}
for some fixed $c \in \RR$. Since $H(0) = 0$ this equation has at
least one solution, $x_0 = 0$. If $r_t \geq \binv$ then $H(x)$ is
strictly concave on $[0,\infty)$, and by Lemma~\ref{Lem:convex}
the equation \eqref{Eq:H_affine} has at most one additional
solution $x_1$. Also, when the solution exists, $H(x)$ crosses
$cx$ from above at $x_1$. Similarly if $r_t \leq \bnorm$ then
$H(x)$ is strictly convex, and there exists at most one additional
solution $x_2$ to \eqref{Eq:H_affine} on $[0,\infty)$. If the
solution exists, then $cx$ is crossed from below at $x_2$. In the
last case $\bnorm < r_t < \binv$, there exists a $x_*$ -- the zero
of $k(x)$ -- such that $H(x)$ is strictly convex on $(0,x_*)$ and
strictly concave on $(x_*,\infty)$. Now there can exist at most
two additional solutions $x_1, x_2$ to \eqref{Eq:H_affine} with
$x_1 < x^* < x_2$, such that $cx$ is crossed from below at $x_1$
and from above at $x_2$.\\
Because of definition \eqref{Eq:H_def}, every solution to
\eqref{Eq:H_affine}, excluding $x_0 = 0$, is also a solution to
\begin{equation}\label{Eq:Y_horizontal}
Y(r_t,x) = c, \qquad x \in (0,\infty)\;
\end{equation}
with $r_t$ fixed. Also the properties of crossing from above/below
are preserved since $x$ is positive. This means that in the case
$r_t \geq \binv$, equation \eqref{Eq:Y_horizontal} has at most a
single solution, or in other words, that every horizontal line is
crossed by the yield curve at most in a single point. If it is
crossed, it is crossed from above. This implies that $Y(x)$ is a
strictly decreasing function of $x$, or following
Definition~\ref{Defn:classify}, that the yield curve is inverse.
In the case $r_t \leq \bnorm$ we have again that
\eqref{Eq:Y_horizontal} has at most a single solution and that
every horizontal line is crossed from below by the yield curve, if
it is crossed. In other words, the yield curve is normal. In the
last case of $\bnorm < r_t < \binv$, the yield curve crosses every
horizontal line at most twice, in which case it crosses first from
below, then from above. Thus in this case the yield curve is
humped.
\end{proof}

\begin{cor}Under the conditions of Theorem \ref{Thm:yield_curves} the instantaneous forward rate curve has
the same global behavior as the yield curve, i.e.
\begin{align*}
Y(r_t,.) \; \text{is inverse} \quad &\Longleftrightarrow \quad f(r_t,.)\;\text{is strictly decreasing}\\
Y(r_t,.) \; \text{is humped} \quad &\Longleftrightarrow \quad \begin{array}{l}f(r_t,.)\;\text{has exactly one local maximum}\\ \text{and no local minimum}\end{array}\\
Y(r_t,.) \; \text{is normal} \quad &\Longleftrightarrow \quad
f(r_t,.)\;\text{is strictly increasing}\;.
\end{align*}
In the second case the maximum of the forward rate curve is
$f(r_t,x_*)$, where $x_*$ solves
\begin{equation}\label{Eq:Cond_fw_rate_max}
r_t = -\frac{F'(B(x))}{R'(B(x))},\quad x \in (0,\infty)\;.
\end{equation}

\end{cor}
\begin{proof}
This follows from the fact that $\partial_x H(x)$ as given in
\eqref{Eq:Dx} is exactly the forward rate $f(r_t,x)$. The
derivative of the forward rate is therefore $\partial_{xx} H(x)$,
which is given in \eqref{Eq:Dxx} as
\[\partial_x f(r_t,x) = \partial_{xx} H(x) = -B'(x)\cdot k(x)\;.\]
The factor $-B'(x) \neq 0$ is always positive, and the possible
sign changes and zeroes of $k(x)$ are discussed in the proof of
Theorem~\ref{Thm:yield_curves}, leading to the stated
equivalences. Equation \eqref{Eq:Cond_fw_rate_max} is simply the
condition $k(x_*) = 0$.
\end{proof}

\begin{cor}\label{Cor:b_nesting}
Under the conditions of Theorem \ref{Thm:yield_curves} it holds
that
\begin{equation}\label{Eq:b_nesting}
\bnorm < \basymp < \binv\;
\end{equation}
whenever the quantities are finite. In addition it holds that
\begin{equation}\label{Eq:bnorm_finite}
D \cap (\bnorm,\binv) \neq \emptyset\;.
\end{equation}
\end{cor}
\begin{rem}
Note that equation \eqref{Eq:bnorm_finite} implies that there is
always some $r_t \in D$ such that the yield curve $Y(r_t,.)$ is
humped.
\end{rem}

\begin{proof}
By the mean value theorem there exists a $\xi \in (-1/\lambda,0)$
such that
\[\basymp = -F(-1/\lambda) = F(0) - F(-1/\lambda) = \frac{1}{\lambda}F'(\xi)\;.\]
Since $F$ is convex and thus $F'$ increasing, it holds that
\begin{equation}\label{Eq:F_sandwich}
\frac{F'(-1/\lambda)}{\lambda} \leq \basymp \leq
\frac{F'(0)}{\lambda}\;.
\end{equation}
Applying the mean value theorem to $R$, there exists another $\xi
\in (-1/\lambda,0)$ such that
\[1 = R(-1/\lambda) - R(0) = - \frac{1}{\lambda}R'(\xi)\;.\]
Since $R'$ is increasing we deduce that $R'(-1/\lambda) \le
-\lambda < 0$. Assuming that also $R'(0) < 0$ we get
\begin{equation}\label{Eq:R_sandwich}
-\frac{1}{R'(-1/\lambda)} \leq \frac{1}{\lambda} \leq
-\frac{1}{R'(0)}\;.
\end{equation}
Since either $F$ or $R$ is non-linear, one of the functions is
strictly convex by Lemma~\ref{Lem:FR_properties}. Consequently
either both inequalities in \eqref{Eq:F_sandwich} or in
\eqref{Eq:R_sandwich} are strict. Putting them together we get
\[-\frac{F'(-1/\lambda)}{R'(-1/\lambda)} < \basymp <
-\frac{F'(0)}{R'(0)}\;,
\]
proving \eqref{Eq:b_nesting} under the assumption that $R'(0) < 0$.\\
If $R'(0) \geq 0$ then by definition $\binv = \infty$. Equation
\eqref{Eq:F_sandwich} still holds, but in \eqref{Eq:R_sandwich}
only the left inequality sign remains valid. Together this still
proves that $\bnorm < \basymp$ and we have shown
\eqref{Eq:b_nesting}.\\

To prove \eqref{Eq:bnorm_finite} we distinguish two cases:
\begin{enumerate}[(a)]
\item $D = \RR$. In this case it is sufficient to prove $-\infty <
\binv$ and $\bnorm < \infty$. Consider first $\binv$. If $R'(0)
\ge 0$ then by definition $\binv = \infty$ and nothing is to
prove. If $R'(0) < 0$ then $\binv = - F'(0) / R'(0)$. By convexity
$F'(0) > -\infty$ and the assertion follows. Consider now $\bnorm
= -F'(-1/\lambda)/R'(-1/\lambda)$. From \eqref{Eq:R_sandwich} we
know that $R'(-1/\lambda) \le -\lambda < 0$. By convexity
$F'(-1/\lambda) < \infty$ and it follows that $\bnorm < \infty$.
\item $D = \Rplus$. In this case it is sufficient to prove $0 \le
\bnorm$ and to apply \eqref{Eq:b_nesting}. As above we have that
$\bnorm = -F'(-1/\lambda)/R'(-1/\lambda)$ and that $R'(-1/\lambda)
\le -\lambda < 0$. By Definition \ref{Def:Admissible}
\[F'(-1/\lambda) = b + \int_{(0,\infty)}{\xi e^{-\xi/\lambda}\,m(d\xi)}\]
with $b \ge 0$. It follows that $F'(-1/\lambda) \ge 0$, proving
the assertion. \qedhere
\end{enumerate}
\end{proof}

The last Corollary of this section shows the interesting fact that
the occurrence of a humped yield curve is a necessary and
sufficient sign of randomness in the short rate model:

\begin{cor}Let the risk-neutral short rate process be given by a one-dimensional affine process $\rt$ satisfying Condition \ref{Cond:rt} with $F \neq 0$ and quasi-mean-reversion $\lambda > 0$.
Then the following statements are equivalent:
\begin{enumerate}[(i)]
    \item There exists a $r_t \in D$ such that $Y(r_t,.)$ is flat.
    \item There exists no $r_t \in D$ such that $Y(r_t,.)$ is
    humped.
    \item The short rate process $\rt$ is deterministic.
    \item $F(u) = b u$ and $R(u) = \beta u$.
\end{enumerate}
\end{cor}

\begin{proof}
Theorem~\ref{Thm:yield_curves}, together with
Corollary~\ref{Cor:b_nesting}, shows already that $\neg (iv)$
implies $\neg (i)$ and $\neg (ii)$. Also, from the form of the
generator in \eqref{Eq:generator_affine}, it is seen that $(iii)$
and $(iv)$ are equivalent. It remains to show that $(iv)$ implies
$(i)$ and $(ii)$. Proceeding as in the proof of
Theorem~\ref{Thm:yield_curves}  we obtain instead of
\eqref{Eq:kx_definition} simply
\[k(x) = b + r_t \beta\;.\]
The yield curve will be humped if and only if $k$ has a single
(isolated) zero in $[0,\infty)$. Since $k$ is a constant function,
this cannot be the case for any $r_t \in D$ and we have shown
$(ii)$. By the same arguments as in the proof of
Theorem~\ref{Thm:yield_curves} the yield curve is flat if and only
if $k$ is constant and equal to $0$. This is the case if $r_t =
-\frac{b}{\beta}$. It remains to show that $r_t \in D$. Note that
$\beta = -\lambda < 0$. In particular $\beta \neq 0$, such that
for $D = \RR$ we are already done. If $D = \Rplus$ we have by the
admissibility conditions in Definition \ref{Def:Admissible} that
$b \ge 0$. Thus $r_t = -\frac{b}{\beta} \ge 0$ and we have shown
$(i)$.
\end{proof}

\subsection{The Limit Distribution of an Affine
Process}\label{Sec:limit} It is well-known that the Gaussian
Ornstein-Uhlenbeck process, for example, converges in law to a
limit distribution and that this distribution is Gaussian. The
goal of this section is to establish a corresponding result for
affine processes. While calculating the marginal distributions of
an affine process involves solving the generalized Riccati
equations \eqref{Eq:Riccati_general}, it will be seen that the
limit distribution is much easier obtained and can be determined
directly from the functions
$F$ and $R$.\\
In the interest rate model considered in the preceding section,
the short rate follows an affine process under the martingale
measure, such that the results will allow us to characterize the
risk-neutral asymptotic short rate distribution. Often also the
limit distribution under the objective measure is of interest, but
the affine property is in general not preserved by an equivalent
change of measure, such that the results are not directly
applicable. Nevertheless, for the sake of tractability, conditions
on the measure change can be imposed, such that the model is
affine under both the objective and the risk-neutral measure. (See
\citet{NicolatoVenardos} for an example from option pricing and
\citet{Cheridito} for more general results). In such a setting the
results can also be applied under the objective
measure.\\

Before we state the result, we want to recall that a real-valued
random variable $L$ is called self-decomposable if for every $c
\in (0,1)$ there exists a random variable $L_c$, independent of
$L$, such that
\[L = cL + L_c \quad \text{for all} \quad c \in (0,1)\;.\]
Since self-decomposability is a distributional property, we will
identify $L$ and its law, and refer to both as
self-decomposable.\\

For OU-type processes, limit distributions have been studied for
some time; the first results can be found in \citet{JurekVervaat}
and \citet{SatoYamazato}. The next theorem summarizes these
results, and can be found in similar form in
\citet[Theorem~17.5]{Sato}:

\begin{thm} \label{Thm:limit_OU}
Let $\rt$ be a OU-type process on $\RR$. If
\begin{equation*}\label{Eq:limit_cond_OU}
\beta < 0 \quad \text{and} \quad \int_{|\xi| > 1}\log |\xi|
\,m(d\xi) < \infty
\end{equation*}
then $\rt$ converges in law to a limit distribution $L$ which is
independent of $r_0$ and has the following properties:
\begin{enumerate}[(i)]
\item $L$ is self-decomposable. \item The cumulant generating
function $\kappa(u) = \log \int_{\RR}{e^{ux}\,dL(x)}$ satisfies
\begin{equation}\label{Eq:Jurek_Vervaat}
\kappa(iu) = \frac{1}{\beta}\int_u^0{\frac{F(is)}{s}ds} \quad
\text{for all} \quad u \in \RR \;.
\end{equation}
\end{enumerate}
Conversely, if $L$ is a self-decomposable distribution on $\RR$
and $\beta < 0$, there exists a unique triplet $(a,b,m)$
satisfying the admissibility conditions of
Definition~\ref{Def:Admissible}, such that $L$ is the limit
distribution of the affine process (of OU-type) given by the
parameters $(a,b,m,\beta)$.
\end{thm}

As discussed in Section~\ref{Sec:preliminaries}, every regular
affine process with state space $\RR$ is of OU-type, such that the
above theorem applies. We now state our corresponding result for
affine processes on $\Rplus$:
\begin{thm}\label{Thm:limit_dis}
Let $\rt$ be a one-dimensional, regular, conservative affine
process with state space $\Rplus$. If
\begin{equation*}\label{Eq:limit_cond}
R'(0) < 0 \quad \text{and} \quad \int_{\xi > 1}\log \xi \,m(d\xi)
< \infty
\end{equation*}
then $\rt$ converges in law to a limit distribution $L$ which is
independent of $r_0$, and whose cumulant generating function
$\kappa$ is given by
\begin{equation}\label{Eq:Gen_Jurek_Vervaat}
\kappa(u) = \int_u^0{\frac{F(s)}{R(s)}ds} \quad \text{for all}
\quad u \in (-\infty,0]\;.
\end{equation}
\end{thm}

\begin{proof}
By Theorem~\ref{Thm:Schachermayer_1} the transition kernel
$p_t(x,.)$ of the process $(r_t)_{t \geq 0}$ has the
characteristic function
\begin{equation*}\label{Eq:kernel_repeat}
\widehat{p}_t(x,u) = \exp\left(\phi(t,u)  + x \psi(t,u)\right)
\end{equation*}
where $\phi$ and $\psi$ satisfy the generalized Riccati equations
\eqref{Eq:Riccati_general} for all $u \in \cU$, and thus in
particular for all $u \in (-\infty,0]$. Since $R(0) = 0$, $0$ is a
critical point of the autonomous ODE \eqref{Eq:Riccati_general_2},
and by the assumption $R'(0) < 0$ it is asymptotically stable. By
the convexity of $R$, $R'(0) < 0$ also implies that $R(u) > 0$ for
all $u \in (-\infty,0)$, such that $\psi(t,u)$ is a strictly
increasing function in $t$ for all $u \in (-\infty,0)$. Since $0$
is the only critical point of \eqref{Eq:Riccati_general_2} on
$(-\infty,0]$ it also follows that
\[\lim_{t \to \infty}\psi(t,u) = 0 \qquad \text{for all} \quad u \in (-\infty,0]\;.\]
Consequently,
\begin{equation}\label{Eq:limit_dis_interm}
\lim_{t \to \infty}\log \widehat{p}_t(x,u) = \lim_{t \to \infty}
\phi(t,u) = \int_0^\infty{F(\psi(r,u))\,dr} =
\int_u^0{\frac{F(s)}{R(s)}\,ds}
\end{equation}
where the last two equalities follow from
\eqref{Eq:Riccati_general} and the transformation $s = \psi(r,u)$.
We will now show that the last integral in
\eqref{Eq:limit_dis_interm} converges absolutely for all $u \in
(-\infty,0]$: Since $R(u) \ge 0$ and $F(u) \le 0$ for all $u \in
(-\infty,0]$ we obtain
\[\int_u^0{\left|\frac{F(s)}{R(s)}\right|\,ds} = - \int_u^0{\frac{F(s)}{R(s)}}\,ds \le -\frac{1}{R'(0)}\int_u^0{\frac{F(s)}{s}\,ds}, \quad u \in (-\infty,0]\;,\]
where the inequality follows from the fact that the convex
function $R$ is supported by its tangent at $0$. From the
definition of $F(u)$ in \eqref{Eq:F_def} it is clear that the
convergence of the last integral will depend only on the jump part
of $F$, i.e. the integral converges if and only if
\begin{equation}\label{Eq:intergral_interm}
    \int_u^0{\frac{1}{s}\int_{(0,\infty)}{\left(e^{s\xi}-1\right)}\,m(d\xi)\,ds}
    < \infty, \quad \text{for all $u \in (-\infty,0]$.}
\end{equation}
Define $M(u,\xi) = \int_u^0\frac{e^{s\xi}-1}{s}ds$. For a fixed $u
\in (-\infty,0]$, it is easily verified that $M(u,\xi) =
\mathcal{O}(\xi)$ as $\xi \to 0$, and that $M(u,\xi) =
\mathcal{O}(\log \xi)$ as $\xi \to \infty$. Since the L\'{e}vy
measure $m(d\xi)$ integrates $(\xi \wedge 1)$ by
Definition~\ref{Def:Admissible}, and $\log \xi \cdot \Ind{\xi
> 1}$ by assumption, it must also integrate $M(u,\xi)$. Applying
Fubini's theorem, \eqref{Eq:intergral_interm} follows, such that
$\kappa(u) := \int_u^0{\frac{F(s)}{R(s)}}\,ds$ converges for all
$u \in (-\infty,0]$. In particular $\lim_{u \uparrow 0}\kappa(u) =
0$, such that the limit in \eqref{Eq:limit_dis_interm} is a
function that is left-continuous at $0$. By standard results on
Laplace transforms of probability measures (cf.
\citet[Theorem~A.3.1]{SteutelHarn}), the pointwise convergence of
cumulant generating functions to a function that is
left-continuous at $0$ implies convergence in distribution of
$\rt$ to a limit distribution $L$ with cumulant generating
function given by \eqref{Eq:limit_dis_interm}.
\end{proof}

Since the marginal distributions of an affine process are
infinitely divisible, also the limit distribution $L$ must be
infinitely divisible, if it exists. In Theorem~\ref{Thm:limit_OU}
a stronger result is given for an affine process on $\RR$: In this
case $L$ is also self-decomposable. An obvious question is, if
this result can be extended to the state space $\Rplus$. We will
see that the answer is negative. In Section~\ref{Sec:JCIR} an
example of an affine process with state space $\Rplus$ is given,
which converges to an infinitely divisible limit distribution that
is not self-decomposable. This result is interesting, since it
leaves open the possibility of some unexpected properties of the
limit distribution of an affine process. For example a
self-decomposable distribution is always unimodal, whereas an
infinitely divisible distribution might be not.

\section{Applications}\label{Sec:applications}

\subsection{The Vasi\v{c}ek model}

We apply the results of the last section to the classical
Vasi\v{c}ek model
\begin{equation}\label{Eq:classical_Vasicek}
dr_t = -\lambda(r_t - \theta)\,dt + \sigma\,dW_t,\quad r_0 \in \RR
\end{equation}
where $(W_t)_{t \ge 0}$ is a standard Brownian motion under the
risk-neutral measure and $\lambda, \theta, \sigma > 0$. The
Vasi\v{c}ek model is arguably the simplest affine model, and no
surprises are to be expected here. In fact all results that we
state here can already be found in the original paper of
\citet{Vasicek}. We advise the reader to view this paragraph as a
warm-up for the
following examples.\\
Clearly $\rt$ is a conservative affine process with
\begin{align}
F(u) &= \lambda \theta u + \frac{\sigma^2}{2}u^2\;,\\
R(u) &= - \lambda u\;.
\end{align}
From the quadratic term in $F$ and
Definition~\ref{Def:Admissible}, it is seen that $\rt$ has state
space $\RR$. This property is often criticized, since it allows
the short rate
to become negative.\\
From Theorem~\ref{Thm:yield_curves} we calculate
\[\binv = \theta \qquad \text{and} \qquad \bnorm = \theta - \frac{\sigma^2}{\lambda^2}\;,\]
such that the yield curve in the Vasi\v{c}ek model is normal if
$r_t \leq \theta - \sigma^2/\lambda^2$, inverse if $r_t \geq
\theta$ and
humped in the remaining cases.\\
The long term yield is calculated from \eqref{Thm:basymp} as
\[\basymp = -F(-1/\lambda) = \theta - \frac{\sigma^2}{2 \lambda^2}\;,\]
in this case exactly the \emph{arithmetic} mean of $b_\text{inv}$
and
$b_\text{norm}$.\\
Theorem~\ref{Thm:limit_OU} applies and the cumulant generating
function $\kappa$ of the risk-neutral limit distribution $L$
satisfies
\[ \kappa(iu) = -\frac{1}{\lambda}\int_u^0{\frac{F(is)}{s}\,ds} =
\int_0^u{\left(i \theta - \frac{\sigma^2}{2 \lambda}s\right)\,ds}
= u i \theta - \frac{u^2}{2}\frac{\sigma^2}{2\lambda}
\]
for $u \in \RR$. Hence, $L$ is Gaussian with mean $\theta$ and
variance
$\frac{\sigma^2}{2\lambda}$.\\

\subsection{The Cox-Ingersoll-Ross model}\label{Sec:CIR}
The Cox-Ingersoll-Ross (CIR)-model was introduced by \citet{CIR}.
In this model the short rate process $\rt$ is given by the SDE
\begin{equation}\label{Eq:CIR}
dr_t = -a(r_t - \theta)dt + \sigma\sqrt{r_t}\,dW_t, \qquad r_0 \in
\Rplus
\end{equation}
where $(W_t)_{t \ge 0}$ is a standard Brownian Motion under the
risk-neutral measure and $a, \theta, \sigma
> 0$. The process $\rt$ is a conservative affine process with
\begin{align}\label{Eq:CIR_FR}
F(u) &= a\theta u\;,\\
R(u) &= \frac{\sigma^2}{2}u^2 - a u\;.
\end{align}
From Definition~\ref{Def:Admissible} it is seen that $\rt$ has
state space $\Rplus$. The fact that interest rates stay
non-negative in the CIR-model is often cited as an advantage of
the model over the Vasi\v{c}ek model. Calculating the
quasi-mean-reversion (see
Definition~\ref{Def:quasi_mean_reversion}), we find that
\begin{equation*}\label{Eq:quasi_mean_rev_CIR}
\lambda = \frac{1}{2}\left(\sqrt{a^2 + 2\sigma^2} + a\right)\;.
\end{equation*}
From Theorem~\ref{Thm:basymp} we find that the long-term yield is
given by
\begin{equation*}\label{Eq:basymp_CIR}
\basymp = -F(-1/\lambda) = \frac{2a\theta}{\sqrt{a^2 + 2\sigma^2}
+ a}\;.
\end{equation*}
The boundary between humped and inverse behavior $\binv$ is
calculated from Theorem~\ref{Thm:yield_curves} as
\begin{equation*}\label{Eq:binv_CIR}
\binv = -\frac{F'(0)}{R'(0)} = \theta.
\end{equation*}
Both quantities $\basymp$ and $\binv$ can also be found in
\citep[Eq.~(26) and following paragraph]{CIR}. Before we consider
$\bnorm$, we quote (with notation adapted to \eqref{Eq:CIR}) from
page 394 of \citep{CIR} where the shape of the yield curve is
discussed:
\begin{quote}
\textit{`When the spot rate is below the long-term yield $[=~
\basymp]$, the term structure is uniformly rising. With an
interest rate in excess of $\theta$ $[=~\binv]$, the term
structure is falling. For intermediate values of the interest
rate, the yield curve is humped.'}
\end{quote}
In our terminology, they claim that the yield curve is normal for
$r_t \leq \basymp$, humped for $\basymp < r_t < \binv$ and inverse
for $r_t \geq \binv$. This stands in clear contradiction to
Theorem~\ref{Thm:yield_curves} and Corollary~\ref{Cor:b_nesting}
where we have obtained that yield curves are normal if and only if
$r_t \leq \bnorm$ and that $\bnorm < \basymp$, or -- in plain
words -- that there are yield curves starting strictly
\emph{below} the
long-term yield that are still humped.\\
The claims of \citet{CIR} are repeated in
\citep[p.~244f]{Rebonato}, where even several plots of `yield
surfaces' (the yield as a function of $r_t$ and $x$) are presented
as evidence. However \citeauthor{Rebonato} fails to indicate the
level of $\basymp$ in the plots, such that the conclusion remains
ambiguous.\\
To clarify the scope of humped yield curves in the CIR-model we
calculate $\bnorm$ from Theorem~\ref{Thm:yield_curves}:

\begin{equation*}\label{Eq:bnorm_CIR}
\bnorm = -\frac{F'(-1/\lambda)}{R'(-1/\lambda)} =
\frac{a\theta}{\sqrt{a^2 + 2\sigma^2}}\;.
\end{equation*}
The relation $\bnorm < \basymp < \binv$ is immediately confirmed
by noting that $\basymp$ is the \emph{harmonic} mean of $\bnorm$
and $\binv$. For a graphical illustration we refer to the second
yield curve from below in Figure~\ref{Fig:yield}. The plot
actually shows CIR yield curves with parameters
\[a = 0.5,\quad \sigma = 0.5,\quad \theta = 6\% \]
plotted over a time scale of $25$ years. The second curve from below starts at $r_0 = 4.2\%$, i.e. below the long-term yield, but is visibly humped.\\

To calculate the limit distribution of $\rt$, we apply
Theorem~\ref{Thm:limit_dis}: The cumulant generating function
$\kappa(u)$ of the limit distribution is given by
\begin{equation*}
\kappa(u) = \int_u^0{\frac{F(s)}{R(s)}\,ds} =
\int_0^u{\frac{\theta}{1 - s\sigma^2/2a}\,ds} =
{-\dfrac{2a\theta}{\sigma^2}} \,\log\left(1 -
\frac{\sigma^2}{2a}u\right)\;.
\end{equation*}
This is the cumulant generating function of a gamma distribution
with shape parameter $2a\theta/\sigma^2$ and scale parameter
$\sigma^2/2a$. Again this result can already be found in
\citet[p.~392]{CIR}.

\subsection{An extension of the CIR model}\label{Sec:JCIR}

To illustrate the power of the affine setting, we consider now an
extension of the CIR model that is obtained by adding jumps to
\eqref{Eq:CIR}. We define the risk-neutral short rate process by
\begin{equation}\label{Eq:CIR_jump}
dr_t = -a(r_t - \theta)dt + \sigma\sqrt{r_t}\,dW_t + dJ_t, \qquad
r_0 \ge 0
\end{equation}
where $(J_t)_{t \geq 0}$ is a compound Poisson process with
intensity $c > 0$ and exponentially distributed jumps of mean $\nu
> 0$. This model has been introduced by \citet{DuffieGarleanu} as a
model for default intensity and is used by \citet{Filipovic} as a
short rate model. It can also be found in \citet{BrigoMerc} under
the name JCIR model. It is easily calculated that
\begin{align}\label{Eq:CIR_jump_FR}
F(u) &= a\theta u + \frac{cu}{\nu - u}, \quad u \in (-\infty,\nu)\;,\\
R(u) &= \frac{\sigma^2}{2}u^2 - a u\;.
\end{align}
Solving the generalized Riccati equations
\eqref{Eq:Riccati_bondprice} for $A(x)$ and $B(x)$ becomes quite
tedious, but the quantities $\binv, \basymp, \bnorm$ can be
calculated from Theorem~\ref{Thm:basymp} and
Theorem~\ref{Thm:yield_curves} in a few lines: The quasi-mean
reversion $\lambda$ stays the same as in the CIR model, since $R$
does not change. From
\[F'(u) = a\theta + \frac{c\nu}{(\nu - u)^2}\]
we derive immediately
\begin{align*}
\binv &= \theta + \frac{c}{a\nu}\;,\\
\basymp &= \frac{2a\theta}{a + \gamma} + \frac{2c}{\nu(a + \nu) +
2}\;,\\
\bnorm &= \frac{a\theta}{\gamma} +
\frac{c\nu\sigma^4}{\gamma(\sigma^2 \nu + \gamma - a)^2}\;,
\end{align*}
where $\gamma = \sqrt{a^2 + 2\sigma^2}$. Note that by setting the
jump intensity $c$ to zero, the expressions
of the (original) CIR model are recovered.\\

Next we calculate the limit distribution of the model. Using the
abbreviations $\rho := \sigma^2/2$ and $\Delta := a - \nu \rho$ we
obtain
\begin{align*}
\kappa(u) &= \int_u^0{\frac{F(s)}{R(s)}\,ds} =
\int_0^u{\frac{\theta}{1 - s\rho/a}\,ds} + c \int_0^u{\frac{ds}{(s
- \nu)(\rho s - a)}} = \\
& = \begin{cases} \left(\frac{c}{\Delta} -
\frac{a\theta}{\rho}\right)\log\left(1 - \frac{\rho}{a}u\right) -
\frac{c}{\Delta}\log\left(1 - \frac{u}{\nu}\right) \quad
&\text{if}
\quad \Delta \neq 0\\
-\theta \nu \log\left(1 - \frac{u}{\nu}\right) + \frac{c}{a}
\frac{u}{\nu - u} \quad &\text{if} \quad \Delta = 0
\end{cases}
\end{align*}
as the cumulant generating function of the limit distribution
$L$ under the martingale measure.\\

We now take a closer look at the distribution $L$, since this will
answer the question raised at the end of Section \ref{Sec:limit}:
For certain parameters, $L$ is an example for a limit distribution
of an affine process that is infinitely divisible, but not
self-decomposable. We consider the case $\Delta = 0$ and define
\begin{equation}\label{Eq:lx_def}
l(x) :=  \left(\theta + \frac{c}{a}x\right) \nu e^{-\nu x}, \qquad
x \in \Rplus\,.
\end{equation}
By Frullani's integral formula
\begin{equation}\label{Eq:JCIR_levy}
\kappa(u) = \int_0^\infty{\left(e^{ux}-1\right)\frac{l(x)}{x}\,dx}
\end{equation}
for all $u \in (-\infty,\nu)$. Since $l$ is non-negative on
$\Rplus$, $l(x)/x$ is the density of a L\'{e}vy measure and
\eqref{Eq:JCIR_levy} is seen to be the L\'{e}vy-Khintchine
representation for the cumulant generating function of the
infinitely divisible distribution $L$. In addition, $L$ is
self-decomposable if and only if $l$ is non-negative and
\emph{non-increasing} on $\Rplus$ (cf.
\citet[Corollary~15.11]{Sato}).\\
In the case of $l(x)$ given by \eqref{Eq:lx_def}, it is easily
calculated that $l(x)$ has a single maximum at $x^* =
\frac{1}{\nu} - \frac{a \theta}{c}$. Thus, if $c \le a \theta
\nu$, then $x^* \le 0$, such that $l$ is non-increasing on
$\Rplus$ and $L$ is self-decomposable. If $ c > a \theta \nu$ then
$l$ is increasing in the interval $[0,x^*)$ and the limit
distribution $L$ is infinitely divisible, but not
self-decomposable.

\subsection{The gamma model}\label{Sec:Gamma}
Instead of analyzing the properties of a known model, we will now
follow a different route and construct a model that satisfies some
given properties. We want to construct an affine process on
$\Rplus$ that has the same limit distribution as the CIR model
(i.e. a gamma distribution), but is a process of OU-type. The
second property is equivalent to $R(u) = \beta u$. Considering
Theorem~\ref{Thm:limit_dis}, we know that if we want to obtain a
limit distribution, we need $\beta < 0$. To keep with the notation
of the Vasi\v{c}ek model, we will write $R(u) = - \lambda u$ where
$\lambda > 0$. Now by \eqref{Eq:Gen_Jurek_Vervaat} the cumulant
generating function of the limit distribution is given by
\begin{equation}\label{Eq:gamma_model}
\kappa(u) = \frac{1}{\lambda}\int_0^u{\frac{F(s)}{s}\,ds} \quad
\text{for all} \quad u \in (-\infty,0]\;.
\end{equation}
Let the limit distribution be a gamma distribution with shape
parameter $k
> 0$ and scale parameter $\theta > 0$. Then $\kappa(u) = -k \log(1 - \theta
u)$ and by \eqref{Eq:gamma_model}
\[F(u) = \frac{\lambda \theta k u}{1 - \theta u}\;.\]
Setting $c = \lambda k$ and $\nu = 1/\theta$ it is seen that
$F(u)$ is equal to the last term in \eqref{Eq:CIR_jump_FR}. This
means that the driving L\'{e}vy process of $\rt$ is of the same
kind as the process $(J_t)_{t \ge 0}$ in \eqref{Eq:CIR_jump}, i.e.
$\rt$ is a pure jump OU-type process with exponentially
distributed jump heights of mean $1/\theta$ and
with jump intensity $\lambda k$.\\
We interpret the affine process we have constructed as a
risk-neutral short rate process. It is clear that the bond prices
are of the exponentially-affine form \eqref{Eq:affine_bond_price}.
From the generalized Riccati equation
\eqref{Eq:Riccati_bondprice2} we obtain
\[B(x) = \frac{e^{-\lambda x}-1}{\lambda}\;.\]
From equation \eqref{Eq:Riccati_bondprice1} we calculate
\begin{equation*}
A(x) = \int_0^x{F(B(s))\,ds} = \frac{\lambda
k}{\theta+\lambda}\left( \log(1- \theta B(x)) - \theta x
\right)\;,
\end{equation*}
such that the bond prices are given by

\begin{equation*}
P(t,t+x) = \exp\left\{ -x\frac{\lambda \theta k}{\theta+\lambda}
+r_t B(x) \right\}  (1 - \theta B(x))^{\frac{\lambda
k}{\theta+\lambda}}\;.
\end{equation*}

The global shape of the yield curve is described by the quantities
\[b_\text{inv} = k \theta, \quad \basymp = \frac{k}{1/\theta + 1/\lambda}, \quad b_\text{norm} = \frac{k/\theta}{(1/\theta + 1/\lambda)^2}\]
and it is seen that for the gamma-OU-process $\basymp$ is the
\emph{geometric} average of $b_\text{inv}$ and $b_\text{norm}$.

\section{Conclusions}
In this article we have given, under very general conditions, a
characterization of the yield curve shapes that are attainable in
term structure models where the risk-neutral short rate is given
by a time-homogenous, one-dimensional affine process. Even though
the parameter space for this class of models is
infinite-dimensional, the scope of attainable yield curves is very
narrow, with only three possible global shapes. In addition we
have given conditions under which an affine process converges to a
limit distribution, and we have characterized the limit
distribution in terms of its cumulant generating function,
extending some known results on OU-type processes.\\
The most obvious question for future research is the extension of
these results to multi-factor models. It is evident from numerical
results that in two-factor models yield curves with e.g. a dip, or
also with a dip and a hump, can be obtained. It would be
interesting to see if more complex shapes can also be produced, or
if there are similar limitations as in the single-factor case.
Also, in the one-factor case the dependence of the yield curve
shape on the current short rate is basically described
 by the intervals $D \cap (-\infty,\bnorm],(\bnorm,\binv)$ and
$[\binv, \infty)$. In the two-factor case the partitioning of the
state-space might be more complex, and we expect to see more
interesting transitions between yield curve types. Another aspect
is, that since affine processes as a general framework become
better understood, extensions of classical models e.g. by adding
jumps, like in the JCIR model described in Section \ref{Sec:JCIR},
become more feasible and attractive for applications.

\bibliographystyle{plainnat}
\bibliography{ornsteinuhlenbeck}

\end{document}